# Emotional VR handshake by controlling skin deformation distribution

Shun Watatani[1], Hikaru Nagano[2], Yuichi Tazaki[1], and Yasuyoshi Yokokohji[1]

[1] Graduate School of Engineering. Kobe University, Kobe, Japan

[2] Graduate School of Science and Technology, Kyoto Institute of Technology, Kyoto, Japan

(Email: nagano@kit.ac.jp)

**Abstract ---** Digital communication tools are limited to visual and auditory information and lack non-verbal information such as touch, which is important for communicating intentions and emotions. In order to solve this problem, the use of haptic technology in digital communication is attracting attention. In this study, we constructed a virtual handshake system that can reproduce distributed haptic information using a wearable device that presents skin deformation. Using the system, we experimentally obtained the correspondence between emotions and handshaking behavior, and constructed a demonstration of handshakes that can express differences in emotions based on the experimental results.

## 1 INTRODUCTION

The development of virtual reality applications, text chat, and video calls has made it easy to communicate digitally with agents in virtual environments and with real people in remote environments. However, in most cases, the information conveyed by these tools is limited to visual and auditory information, and no consideration is given to the conveyance of other nonverbal information.

In this study, we focus on tactile information as non-verbal information. In recent years, technologies for transmitting tactile information have attracted attention [1-3]. Non-verbal information such as tactile information can supplement verbal information, and is considered to play an important role in the communication of emotions and intentions [4]. Therefore, it is possible to further improve the experience in digital communication environments by using haptic information transfer technology.

Digital tactile communication reproduces tactile interactions that occur in real environments, such as touch [5], hugging [6], and massage [7]. These tactile interactions are reported to improve the experience of digital communication. However, the application is limited to the current digital communication environment, in which people interact with a variety of people in different relationships.

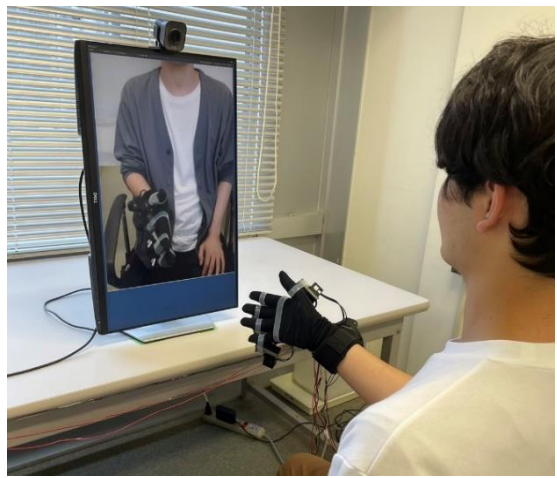

Fig. 1    Demonstration of VR handshake.

The handshake has been attracting attention as a more universal tactile interaction, and various methods have been used to realize a handshake in digital environments. A virtual handshake method that reproduces arm movements [8] and the sensation of a hand being held [9] using a robotic hand has been reported as a representative method. Such handshake reproduction using a robot hand has the advantage of providing a realistic handshake experience. On the other hand, it has the problem that it is strongly restricted by the location, because the devices are large and most of them are installed. Therefore, it is difficult to introduce a method using a robotic hand in a general environment such as a home. In a virtual reality environment, the robot's range of motion may limit the user's movement, making it a poor match.

A virtual handshake technique using a small device that improves accessibility in a variety of situations has also been proposed. As a virtual handshake technique using a wearable device, a skin deformation presentation device using pouch actuator [10] or shape memory alloy [11] has been proposed. Both methods are capable of reproducing a sense of temperature along with skin deformation, but they did not sufficiently consider the intensity of the stimulus. In a real environment, haptic perception typically utilizes spatial distribution information such as stimulus intensity. For instance, the relationship between distribution information and tactile perception has been investigated for the fingertips [12,13] and arms [14]. A handshake also involves spatial distribution information, where different stimuli act on different parts of the hand. Furthermore, the comparative experiments between real and virtual handshakes conducted by Bailenson et al. suggest that stimulus intensity is related to the transmission of emotions [15]. Therefore, this study focuses on the distribution information of stimulation intensity occurring in the hand during handshakes.

In this study, we construct a virtual handshake system that can reproduce distribution information using a wearable device that presents skin deformation. We experimentally obtained the correspondence between emotions and movements, and based on this, we constructed a demonstration as shown in Figure 1 that presents a handshake that can express different emotions.

## 2 METHODS

### 2.1 VR handshake system

The intensity of tactile stimuli produced by shaking hands in a real environment is determined based on the actions of both parties, rather than reflecting the actions of only one party, as is the case with touch. This feature is thought to affect stimulus distribution information. Therefore, in a preliminary experiment, we measured the pressure applied to the hand during an actual handshake and found that the strength of the palm pressure was strongly influenced by the opponent's grip strength, while the finger pressure was strongly influenced by one's own grip strength. Based on this relationship, the system in this study measures the grip strength of both parties during the VR handshake, and reflects the values in the stimulus strength for each area. In this way, the system reflects the movements of the two persons and also expresses information on the distribution of the stimuli. To construct this system, we built a device equipped with a "finger joint angle measurement function" and a "skin deformation presentation function".

### 2.2 Finger joint angle measurement function

A finger joint angle measurement function was implemented to acquire finger flexion motion during handshaking. A flex sensor (SENSIA-BS-65, Sensia Technology Co., Ltd.) attached to a glove is used for the measurement. The two joint positions for angle measurement are the first joint of the thumb and the second joint of the middle finger. The other fingers are not measured because their flexion angles during handshaking are the same as those of the middle finger.

### 2.3 Skin deformation presenting function

When shaking hands, skin deformation stimuli are generated over a wide area of the hand due to contact with the other person's hand. We implemented the function shown in Figure 2, which presents stimuli at multiple points individually, in order to achieve an experience similar to the real environment during a VR handshake. The silicone rubber band is wound around the fingers and palm by a motor, and the skin deformation can be presented to the fingers and palm. Compared to other methods such as vibration presentation, this method can present stimuli that more closely resemble an actual handshake. It also has the feature that the intensity can be adjusted for each part of the body. A small gear motor (Dynamixel XC-330-T288-T, ROBOTIS Co., Ltd.) of $2 \times 3.4 \times 2.6 cm^3$ was used as the motor for winding. Total weight of the device is 250g and power consumption is less than 90W. The following seven stimulus presentation sites were selected based on the results of a study [16] that investigated the contact sites of handshakes.

- Middle tubercles of index, middle, ring and little.
- Base of thumb.
- Lateral surface of palm.

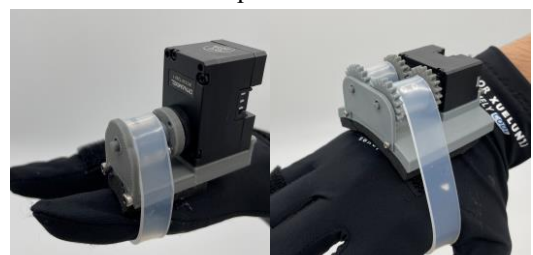

(a) Finger.  (b) Plam.
Fig. 2  Skin deformation presenting function.

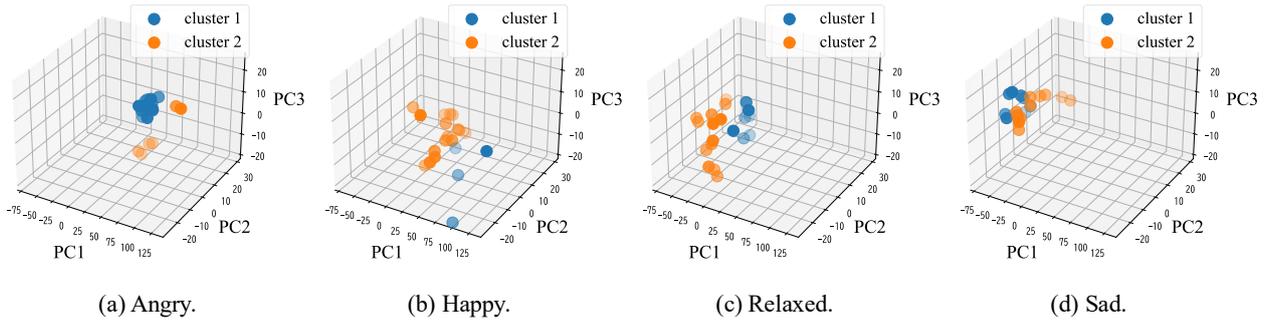

(a) Angry.    (b) Happy.    (c) Relaxed.    (d) Sad.

Fig. 3　The results of principal component analysis and emotion-based clustering conducted on the handshake elements recorded in each dataset.

## 3　EMOTIONAL EXPRESSION VIA VR HANDSHAKE

It is known that there is a correspondence between the parameters of stimuli and the type of emotion in emotional expression via tactile and visual stimuli [15,17]. The investigation of such correspondence can be used to improve the agent's expressive ability and to efficiently select information to be communicated according to the purpose. Therefore, we experimentally obtained the correspondence between emotions and handshake elements in the constructed system as well. In this experiment, we focused on five handshake factors: "grip strength," "grip speed," "swing range," "swing speed," and "handshake duration." These elements are easily adjustable by the participants in terms of emotional expression, assuming that the environment is combined with visual information. In addition, we specified four types of emotions to be expressed, "Angry," "Happy," "Relaxed," and "Sad," based on the circular model of emotion proposed by Russell et al [12].

### 3.1　Experimental device

The finger joint angle measurement function of the developed handshaking system was used to measure "grip strength", "grip speed" and "handshake duration.". Next, a VR tracker (VIVE Tracker 2018, HTC Corp.) attached to the wrist was used for "swing range" and "swing speed". In addition, the skin deformation presentation function included in the remote handshake system was used to provide feedback to participants on the skin deformation stimulus during the handshake.

### 3.2　Experimental procedure

First, to familiarize the participants with the operation of the device, they were allowed to practice operating it freely. Next, the participants were asked to decide on a handshake strategy to express each emotion, based on an explanation of the five factors to be evaluated in this study. The participants then shook hands twice for each emotion using the device, and the handshaking movements were measured. The participants were 10 males in their 20s belonging to our laboratory.

### 3.3　Results: Correspondence map between emotion and behavior

A total of 80 data were obtained by measurement. For each data set, numerical values for the five handshake elements were calculated. Principal component analysis was applied to the calculated values to compress the five handshake elements into three dimensions. The compressed data were then clustered by emotion using the Ward method. The results are shown in the Figure 3.

The plot of each emotion data in the three-dimensional space shows that they are distributed in different regions depending on the emotion. This suggests that there is a general tendency for participants to express their emotions via remote handshakes. Furthermore, clustering revealed that each of the four types of emotional expressions can be further divided into two more detailed tendencies.

## 4　DEMONSTRATION

Using the developed system, participants will experience emotion transmission through a remote handshake. Participants will wear a device equipped with a skin deformation presentation function. Based on the handshake data recorded during the experiment that expresses various emotions, participants will experience different types of handshakes corresponding to the expressed emotions. During this process, recorded videos will be played in sync with the tactile stimuli provided by the device, allowing participants to experience a handshake that combines both visual and tactile sensations.

## 5 Conclusion

To improve the experience of communicating with agents in a virtual environment and with real people in a remote location, we focused on handshaking through the transmission of tactile information. We developed a wearable tactile display that reproduces the tactile distribution during a handshake. We also developed a demonstration of a VR handshake system that expresses differences in emotions by using a map that shows the correspondence between emotions and handshake actions.

### Acknowledgement

This work was supported by JSPS KAKENHI Grant Numbers JP21KK0182, JP24K07405 and JP24K21323.